\documentclass[twocolumn,trackchanges,twocolappendix]{aastex701}
\usepackage{color}
\usepackage{bm}
\usepackage{multirow}
\usepackage{textcomp}
\usepackage{amsmath}%

\def \pccm {\mathrm{pc\ cm^{-3}}}
\def \kmsMpc {\mathrm{km\ s^{-1}\ Mpc^{-1}}}

\begin{document}
	
	\title{Dispersion Measure Distribution of Unlocalized Fast Radio Bursts as a Probe of the Hubble Constant}
	
	\author{Yang Liu}
	\email{yangliu@pmo.ac.cn}
	\affiliation{Purple Mountain Observatory, Chinese Academy of Sciences, No. 10 Yuanhua Road, Nanjing 210023, China}

	\author{Jun-Jie Wei}
	\email{jjwei@pmo.ac.cn}
	\affiliation{Purple Mountain Observatory, Chinese Academy of Sciences, No. 10 Yuanhua Road, Nanjing 210023, China}

    \author{Puxun Wu}
    \email{pxwu@hunnu.edu.cn}
    \affiliation{Department of Physics, Key Laboratory of Low Dimensional Quantum Structures and Quantum Control of Ministry of Education, and Hunan Research Center of the Basic Discipline for Quantum Effects and Quantum Technologies, Hunan Normal University, Changsha, Hunan 410081, China}

	\author{Xue-Feng Wu}
	\email{xfwu@pmo.ac.cn}
	\affiliation{Purple Mountain Observatory, Chinese Academy of Sciences, No. 10 Yuanhua Road, Nanjing 210023, China}

	\correspondingauthor{Jun-Jie Wei, Puxun Wu, Xue-Feng Wu}
	\email{jjwei@pmo.ac.cn, pxwu@hunnu.edu.cn, xfwu@pmo.ac.cn}

\begin{abstract}
We present constraints on the Hubble constant ($H_0$) derived from the observed dispersion measure (DM) distribution of unlocalized fast radio bursts (FRBs). While localized FRBs with redshift measurements have been used to investigate the Hubble tension, their sample remains limited. Here we demonstrate that unlocalized FRBs---which are far more numerous---can independently constrain $H_0$ without requiring redshift information, as cosmic expansion imprints itself on their DM distribution. 
Analyzing a selected sample of 2124 unlocalized FRBs from the CHIME Catalog II, we obtain $H_0 = 69^{+17}_{-15}~\mathrm{km\,s^{-1}\,Mpc^{-1}}$ at the $1\sigma$ confidence level, corresponding to an uncertainty of about 22\%.
Disentangling the parametric degeneracy among $H_0$, the FRB spectral index 
$\alpha$, and the characteristic cutoff energy $E_*$ of the FRB energy distribution would reduce the fractional uncertainty in $H_0$ to 9\%. This work constitutes the first $H_0$ measurement derived solely from the DM distribution of unlocalized FRBs, highlighting their potential as a new cosmological probe. Future joint analyses with localized FRBs promise even tighter constraints.
\end{abstract}


		
\section{Introduction}
The Hubble tension has emerged as one of the most significant crises in modern cosmology, referring primarily to the discrepancy between the value of the Hubble constant ($H_0$) inferred from the cosmic microwave background (CMB) within the $\Lambda$CDM framework and that measured from local distance indicators independent of the cosmological model.
Recently, the H0DN collaboration reported a precise measurement of $H_0=73.50\pm0.81~\kmsMpc$ by combining multiple local distance indicators~\citep{2026A&A...708A.166H}.
This value is significantly higher than the latest CMB-inferred value of $H_0=67.24\pm0.35~\kmsMpc$~\citep{2026PhRvD.113h3504C}, thereby raising the statistical significance of the discrepancy from the previously reported level of about $5\sigma$~\citep{2020A&A...641A...6P,2022ApJ...934L...7R} to $7.1\sigma$.
Given that extensive searches for systematic errors have so far failed to resolve this discrepancy~\citep{2021CQGra..38o3001D,2023ApJ...944...94T,2023ARNPS..73..153K,2024ApJ...966...89A,2025ApJ...992L..34R}, independent probes are essential to determine whether this tension originates from unrecognized systematic errors or from new physics beyond the $\Lambda$CDM model.

Fast Radio Bursts (FRBs) are millisecond-duration astronomical radio transients characterized by extremely high isotropic equivalent energies~\citep{2007Sci...318..777L}. 
Their dispersion measures (DMs) depend on both the source distance and the number density of free electrons along the line of sight.
Extragalactic FRBs therefore serve as a powerful probe for cosmological studies~\citep{2014ApJ...783L..35D, 2014ApJ...788..189G, 2014PhRvD..89j7303Z, 2018ApJ...856...65W,2018ApJ...860L...7W, 2019A&ARv..27....4P, 2022A&ARv..30....2P, 2024ChPhL..41k9801W}. 
In particular, FRBs can provide an independent constraint on $H_0$ that is complementary to both CMB data and local distance indicators, potentially helping to clarify the origin of the Hubble tension~\citep{2018NatCo...9.3833L, 2022MNRAS.511..662H, 2022MNRAS.516.4862J, 2022MNRAS.515L...1W, 2022arXiv221213433Z, 2023ApJ...946L..49L, 2023ApJ...955..101W, 2025A&A...693A..85Y, 2025JCAP...01..018F, 2025PDU....4801926K, 2025A&A...698A.215G, 2025PhRvD.112j3541J, 2025ApJ...981....9W, 2025ApJ...988..177X, 2026ApJ...996...66Z, 2025SCPMA..6880406Z, 2025arXiv250706975S, 2026arXiv260629583R, 2026arXiv260109521W, 2026arXiv260622390W}.

However, the commonly used method for constraining $H_0$ with FRBs relies on the DM--$z$ relation and therefore requires each burst to be localized to its host galaxy and to have a measured redshift $z$. At present, the sample of localized FRBs remains limited to just over a hundred sources. Although this sample already constrains $H_0$ to a precision of about 3\%~\citep{2025ApJ...988..177X}, achieving $\sim1\%$ precision would require a substantially larger sample of approximately 500 localized FRBs~\citep{2025A&A...698A.215G}.
The limited number of localized sources thus hinders further improvement of $H_0$ constraints using FRBs alone, motivating the exploration of methods that do not require redshift measurements. 
{ Several such attempts have been made in previous studies.} 
\cite{2022MNRAS.516.4862J} constrained $H_0$ using a joint analysis of localized and unlocalized FRBs, but their constraint is still driven mainly by surveys containing localized FRBs with measured redshifts.
Inspired by the dark siren method in gravitational-wave cosmology, \cite{2022arXiv221213433Z} developed a Bayesian framework that combines unlocalized FRBs with galaxy catalog data to infer $H_0$. This method depends on external galaxy catalogs, and it is limited by the completeness of the galaxy catalog at higher redshifts.
More recently, \cite{2025A&A...698A.215G} estimated pseudo-redshift distributions for 527 unlocalized FRBs through Monte Carlo sampling and then constrained $H_0$ by incorporating these pseudo-redshift estimates into the FRB analysis. Since the pseudo-redshift estimation itself depends on an assumed value of $H_0$, their analysis requires an iterative treatment to handle the resulting circularity. Moreover, the pseudo-redshift estimates may be affected by DM contributions from the host galaxy or the local environment of the source, especially for low-redshift FRBs~\citep{2023ChPhC..47h5105T}. 
The angular auto-power spectrum of FRB DMs has recently been measured with the latest unlocalized FRB data and used to constrain $H_0$ and the cosmic baryon density, although the current sample is not yet sufficient for a precise $H_0$ constraint~\citep{2026arXiv260704106W}. 
When combined with external data sets, unlocalized FRBs enable further probes of cosmology and baryonic matter, for example through cross-correlations with foreground dark matter halos~\citep{2022MNRAS.512.1730S} or with galaxy catalogs and other large-scale structure tracers~\citep{2020PhRvD.102b3528R, 2021ApJ...922...42R, 2026PhRvL.137d1001W, 2026ApJ...998..109S, 2026arXiv260422105S}, and through joint analyses with gravitational-wave sources~\citep{2026arXiv260403163Z}.
So far, it remains an important question whether $H_0$ can be constrained using only unlocalized FRBs, without relying on redshift measurements or other external observations.
In fact, because cosmic expansion governs the DM--$z$ relation, its imprint must also be encoded in the observed DM distribution of unlocalized FRBs, a population that is orders of magnitude larger than the localized sample.
Consequently, even without individual redshift measurements, the DM distribution of unlocalized FRBs contains valuable cosmological information, as explored in forecast studies of the cosmic helium reionization history~\citep{2021PhRvD.103j3526B}. This offers an alternative route to constrain the Hubble constant.

Motivated by this prospect, we propose in this work to infer $H_0$ using the observed DM distribution of unlocalized FRBs. Applying this method to a selected sample of 2124 unlocalized FRBs from the latest CHIME Catalog II, released by the Canadian Hydrogen Intensity Mapping Experiment (CHIME) collaboration~\citep{2026ApJS..283...34C}, we obtain a constraint on $H_0$ with an uncertainty of about 22\%, conditional on the baryon density $\Omega_{\mathrm{b0}}h^2$ being fixed to the CMB-inferred value. This marks the first $H_0$ constraint derived solely from the DM distribution of unlocalized FRBs.
While the precision of this method is currently lower than that from localized FRBs, it offers several key advantages. It does not rely on host galaxy identification or redshift follow-up, and it can be directly applied to the vast majority of detected FRBs.
Moreover, our results indicate that the precision can be significantly improved with larger samples and better calibration of the FRB energy distribution. Thus, unlocalized FRBs represent a promising complementary probe of the Hubble tension. As a powerful complement to the limited sample of localized FRBs, their abundant population holds promise for enhancing the overall precision of $H_0$ constraints obtained from FRBs.
{Since no host galaxy identification or redshift follow-up is required, the unlocalized sample is expected to grow rapidly with ongoing and future surveys. With CHIME/FRB continuing to operate~\citep{2026ApJS..283...34C} and future survey telescopes such as DSA-2000~\citep{2019BAAS...51g.255H}, the sample required for a percent-level $H_0$ measurement is expected to become available in the near future.}

\section{Methodology}\label{sec:method}
For an extragalactic FRB, the observed dispersion measure ($\mathrm{DM_{obs}}$) can be decomposed into Galactic and extragalactic components:
\begin{equation}\label{eq:DMobs}
	\mathrm{DM_{obs}} = \mathrm{DM_{MW}} + \mathrm{DM_{ext}},
\end{equation}
where $\mathrm{DM_{MW}}$ denotes the Milky Way contribution and $\mathrm{DM_{ext}}$ represents the extragalactic component. 
The Galactic term further splits as $\mathrm{DM_{MW}} = \mathrm{DM_{ISM}} + \mathrm{DM_{halo}}$, with the subscripts ``ISM'' and ``halo'' referring to the Galactic interstellar medium and halo, respectively.
We calculate $\mathrm{DM_{ISM}}$ using the YMW16 model~\citep{2017ApJ...835...29Y} and treat $\mathrm{DM_{halo}}$ as a free parameter in the subsequent analysis.

The component most relevant to cosmology is $\mathrm{DM_{ext}}$, which comprises contributions from the FRB host galaxy ($\mathrm{DM_{host}}$) and from the intergalactic medium (IGM) together with intervening halos ($\mathrm{DM_{cos}}$).
For an FRB with a known redshift, the probability density function (PDF) of $\mathrm{DM_{ext}}$ can be expressed as:
\begin{align}\label{eq:Pext|z}
	&P_\mathrm{ext}(\mathrm{DM_{ext}}|z,\bm{\Theta}) =  \int_{0}^{ \mathrm{DM}_{\mathrm{ext}}(1+z) } 
	P_{\mathrm{host}}(\mathrm{DM}_{\mathrm{host}}|\bm{\Theta}) \nonumber \\
	&\quad\quad \times P_{\mathrm{cos}}( \mathrm{DM}_{\mathrm{ext}} -\mathrm{DM}_{\mathrm{host}}/(1+z)|\bm{\Theta})
	\ d\mathrm{DM}_{\mathrm{host}}, 
\end{align}
where the $P_{\mathrm{host}}$ is the PDF of $\mathrm{DM_{host}}$ in the FRB rest frame. It is commonly modeled using a log-normal distribution~\citep{2020Natur.581..391M}, characterized by the parameters $\mu_\mathrm{host}$ and $\sigma_{\mathrm{host}}$:
\begin{align}\label{eq:P_host}
	P_\mathrm{host}(\mathrm{DM}_\mathrm{host}) &= 
	\frac{1}{\sqrt{2\pi} \, \mathrm{DM}_\mathrm{host} \, \sigma_{\mathrm{host}}} \nonumber \\
	&\times\exp\left(
	-\frac{\left(\ln \mathrm{DM}_\mathrm{host} - \mu_{\mathrm{host}}\right)^2}{2\sigma_{\mathrm{host}}^2}
	\right),
\end{align}
which we treat as free parameters to be constrained jointly with others.
Several parameterizations of $P_{\mathrm{cos}}$ have been proposed in the literature~\citep{2025MNRAS.540..289G, 2025ApJ...989...81S, 2025arXiv250707090K, 2026PhRvD.113d3513Z}. In this work, we adopt the widely used form introduced by~\cite{2020Natur.581..391M}, which is typically expressed as~\citep{2020Natur.581..391M, 2014ApJ...780L..33M, 2025PhRvD.112h3516Z, 2026ApJ...996...66Z}:
\begin{equation} \label{eq:P_cos}
	P_{\mathrm{cos}}(\mathrm{DM_{cos}}|\bm{\Theta}) = \frac{1}{\langle\mathrm{DM_{cos}}\rangle} P_{\mathrm{\Delta}}\left(\frac{\mathrm{DM_{cos}}}{\langle\mathrm{DM_{cos}}\rangle}  \right),
\end{equation}
where $\Delta \equiv \mathrm{DM_{cos}}/\langle\mathrm{DM_{cos}}\rangle$ denotes the dimensionless
ratio, and its probability density function is given by
\begin{equation}\label{eq:P_Delta}
	P_{\mathrm{\Delta}}(\Delta)=A\Delta^{-\beta}\exp\left[-\frac{(\Delta^{-\alpha}-C_0)^2}{2\alpha^2\sigma_{d}^2}\right],\ \Delta>0.
\end{equation}
Here we adopt $\alpha=\beta=3$, $A$ is a normalization factor, and $C_0$ is chosen such that the mean of $\Delta$ satisfies $\langle \Delta \rangle = 1$. The parameter $\sigma_d$ is related to the standard deviation of $\mathrm{DM_{cos}}$ and is commonly parameterized as $\sigma_d = f z^{-0.5}$, where $f$ characterizes the strength of baryon feedback. 
Following the recent estimate in \citep{2024ApJ...965...57B}, we set $f=10^{-0.75}$ in our analysis.
{The log-normal form of $P_\mathrm{host}$ and the Macquart form of $P_\mathrm{cos}$ adopted above are commonly used in the literature.
However, these functional forms are uncertain given our limited knowledge of $\mathrm{DM_{host}}$ and the scatter of $\mathrm{DM_{cos}}$.
For analyses based on localized FRBs, different choices of the parameters of these distributions, such as $\mu_\mathrm{host}$, $\sigma_\mathrm{host}$, and the parameter $f$, have been shown to shift the inferred $H_0$ at the level of several sigma~\citep{2025ApJ...988..177X}.
In our analysis, $\mu_\mathrm{host}$ and $\sigma_\mathrm{host}$ are kept free, so this source of uncertainty may be largely absorbed in their fit.
In contrast, $f$ is fixed in this work. We test the effect of freeing it in the following analysis, and find that the inferred $H_0$ remains essentially unchanged at the current statistical precision.
As for the functional forms themselves, although some hydrodynamical simulation studies suggest that $P_\mathrm{cos}$ may deviate from the Macquart form~\citep{2025ApJ...989...81S, 2025arXiv250707090K}, our previous analysis found that adopting different parameterizations of $P_\mathrm{cos}$, including the log-normal form, changes the inferred constraints only slightly~\citep{2026PhRvD.114d3537L}.
However, that test was based on localized FRBs and covered only a limited set of forms. 
Whether the adopted functional forms of $P_\mathrm{cos}$ and $P_\mathrm{host}$ significantly affect the $H_0$ constraint from the $\mathrm{DM_{obs}}$ distribution of unlocalized FRBs is beyond the scope of this work and is left to future work.
}

The mean value $\langle\mathrm{DM_{cos}}\rangle$ in Eq.~(\ref{eq:P_cos}) depends directly on the cosmological model and can be computed as~\citep{2003ApJ...598L..79I, 2004MNRAS.348..999I, 2014ApJ...783L..35D, 2020Natur.581..391M}:
\begin{align}\label{eq:DMcos}
	\langle\mathrm{DM}_{\mathrm{cos}}\rangle(z)&=\frac{3c\ \Omega_\mathrm{b0}h^2(100~\kmsMpc)^2}{8\pi G m_p H_0} \nonumber \\
	&\times \int_{0}^{z}\frac{(1 + z')\ f_{d}\ \chi_{e}(z')}{\sqrt{\Omega_\mathrm{m0}(1 + z')^{3}+ \left(1 - \Omega_\mathrm{m0}\right) }}\mathrm{d}z'.
\end{align}
Here $c$, $G$, $m_p$, $\Omega_\mathrm{m0}$, and $H_0$ are the speed of light, the gravitational constant, the proton mass, the present matter density parameter, and the Hubble constant, respectively. The parameter $\Omega_\mathrm{b0}h^2$ (with $h\equiv H_0/(100~\kmsMpc)$) is the current physical baryon density. In this equation, we have assumed a spatially flat $\Lambda$CDM model. 
The parameter $f_d$ quantifies the fraction of cosmic baryons residing in the diffuse gas (i.e., IGM and extragalactic halos), whereas $\chi_e(z)$ describes the ionization state as the effective number of free electrons per baryon:
$\chi_e(z)=Y_\mathrm{H} \chi_{e,\mathrm{H}}(z)+\frac{1}{2}Y_\mathrm{He}\chi_{e,\mathrm{He}}(z)$, with $Y_\mathrm{H}\simeq 3/4$ and $Y_\mathrm{He}\simeq 1/4$ being the hydrogen and helium mass fractions, and $\chi_{e,\mathrm{H}}$, $\chi_{e,\mathrm{He}}$ their ionization fractions. 
Since both hydrogen and helium are fully ionized at $z<3$~\citep{2009RvMP...81.1405M,2011MNRAS.410.1096B}, we take $\chi_{e,\mathrm{H}}=\chi_{e,\mathrm{He}}=1$, yielding $\chi_e=7/8$.

It is straightforward to see that cosmological parameters affect the shape of $P_\mathrm{cos}$, thereby determining the overall form of $P_\mathrm{ext}$ and, in turn, the PDF of the observed $\mathrm{DM_{obs}}$ distribution. Consequently, by modeling the theoretical distribution of $\mathrm{DM_{obs}}$, we can extract cosmological information from unlocalized FRBs.
Since the distribution of $\mathrm{DM_{obs}}$ depends not only on $P_\mathrm{ext}$ but also on the redshift distribution of the FRB population and the telescope detection probability for FRBs at different redshifts, the normalized PDF of $\mathrm{DM_{obs}}$ can be modeled as~\citep{2022MNRAS.509.4775J, 2026ApJ...999..202S}:
\begin{widetext}
	\begin{equation}\label{eq:P_obs}
		P(\mathrm{DM_{obs}}|\bm{\Theta}) = \frac{\int_{0}^{z_\mathrm{max}} P_\mathrm{ext}(\mathrm{DM_{obs}}-\mathrm{DM_{MW}}|z,\bm{\Theta}) P_{z}(z|\bm{\Theta}) P_\mathrm{det}(z|\bm{\Theta}) S_\mathrm{DM}(\mathrm{DM_{obs}}) \, dz}{\int_{0}^{\mathrm{DM_{max}}}\int_{0}^{z_\mathrm{max}} P_\mathrm{ext}(\mathrm{DM_{ext}}|z,\bm{\Theta}) P_{z}(z|\bm{\Theta}) P_\mathrm{det}(z|\bm{\Theta}) S_\mathrm{DM}(\mathrm{DM_{ext}} + \mathrm{DM_{MW}}) \, dz \, d\mathrm{DM_{ext}}}.
	\end{equation}
\end{widetext}
Here $P_{z}$, $P_\mathrm{det}$, and $S_\mathrm{DM}$ denote, respectively, the PDF of the redshift distribution of FRB population, the detected probability as a function of redshift, and the selection function of the $\mathrm{DM_{obs}}$.
We set the integration limits to $\mathrm{DM_{max}} = 4000~\mathrm{pc~cm^{-3}}$ and $z_\mathrm{max}=3$, as the highest redshift of FRBs discovered to date is $z=2.148$~\citep{2025arXiv250801648C}.
The parameter $\bm{\Theta}$ denotes an arbitrary set of parameters.

The term $P_{z}$, which represents the redshift distribution of the FRB population, is given by
\begin{equation}
	P_{z}(z|\bm{\Theta}) \propto \frac{\mathcal{R}(z|\bm{\Theta})}{1+z} \frac{d V_c}{dz}(z|\bm{\Theta}),
\end{equation}
where $dV_c/dz(z|\bm{\Theta})$ is the differential comoving volume. The function $\mathcal{R}(z)$ describes the intrinsic event rate of FRBs, which we parameterize in terms of the cosmic star-formation rate (SFR) density~\citep{2014ARA&A..52..415M}:
\begin{equation}\label{eq:Rz}
	\mathcal{R}(z|\bm{\Theta}) = \left( \frac{\mathrm{SFR}(z)}{\mathrm{SFR}(0)} \right)^n,
\end{equation}
with 
\begin{equation}
	\mathrm{SFR}(z)\propto \frac{(1+z)^{2.7}}{1+\left(\frac{1+z}{2.9}\right)^{5.6}}.
\end{equation}
The free parameter $n$ quantifies possible deviations of the true FRB event rate from the SFR.

The term $P_\mathrm{det}(z|\bm{\Theta})$ represents the probability of detecting an FRB at redshift $z$. 
For simplicity, we assume that it depends only on the intrinsic energy distribution of bursts and on the redshift through the cosmological scaling factor, and we model it as:
\begin{equation}\label{eq:P_det}
	P_\mathrm{det}(z|\bm{\Theta})  = \int_{E_{\mathrm{min}}}^{\infty} \Phi(E) S_F(E/K(z|\bm{\Theta}))\,\mathrm{d}E,
\end{equation}
where $E_{\mathrm{min}}=10^{39}~\mathrm{erg}$ is the adopted minimum isotropic-equivalent energy threshold (below which bursts are assumed to be undetectable). 
The function $\Phi(E)$ is the burst energy distribution. A variety of parametric forms have been explored in the literature~(e.g., \citealt{2018MNRAS.480.4211M, 2018MNRAS.481.2320L, 2019ApJ...883...40L, 2020MNRAS.498.1973L, 2020MNRAS.494..665L, 2021MNRAS.501..157Z, 2022MNRAS.510L..18J, 2022MNRAS.509.4775J, 2022MNRAS.511.1961H, 2023MNRAS.522.3349B, 2023ApJ...944..105S, 2025PASA...42...17H, 2025ApJ...990..175L, 2025PASA...42....3A, 2026MNRAS.545f2148N}). Recent work has shown that a Schechter function provides a good empirical description of the observed energy distribution~(see \citealt{2025ApJ...990..175L}); accordingly, we adopt this form~(originally proposed by \citealt{1976ApJ...203..297S}):
\begin{equation}\label{eq:PhiE}
	\Phi(E)\,{d}E \propto \left( \frac{E}{E_*} \right)^{\gamma} \exp\left( -\frac{E}{E_*} \right) \,{d}\left( \frac{E}{E_*} \right),
\end{equation}
where $E_*$ and $\gamma$ are the characteristic cutoff energy and the power-law index, respectively. These are treated as free parameters to be constrained jointly with the others.
The function $S_F(E/K(z|\bm{\Theta}))$ is the instrumental selection function of CHIME in terms of fluence $F$, which describes the detection probability of an FRB with a given observed $F$ after marginalizing over other burst properties (e.g., scattering time, DM, intrinsic width, spectral index, and spectral running)~\citep{2021ApJS..257...59C}.
Here, we adopt an empirical form motivated by the CHIME Catalog I~\citep{2021ApJS..257...59C}, as given by \citep{2022MNRAS.511.1961H}:
\begin{equation}
	\lg S_F(F) = 1.7173(1-\exp(-2.0348\lg F)) -1.7173,
\end{equation}
where $\lg$ denotes the base-10 logarithm. The conversion factor $K(z|\bm{\Theta})$ between energy $E$ and fluence $F$ is defined by $E = F K(z|\bm{\Theta})$, and is given by
\begin{equation}
	K(z|\bm{\Theta}) = \Delta \nu \frac{4\pi d_L^2(z|\bm{\Theta})}{(1 + z)^{2+\alpha}},
\end{equation}
where $\Delta \nu$ is the frequency bandwidth (taken as $1~\mathrm{GHz}$) 
and $\alpha$ is the FRB spectral index. 
The quantity {$d_L(z|\bm{\Theta})$ is the luminosity distance. Like $dV_c/dz(z|\bm{\Theta})$, it varies with the cosmological parameters contained in $\bm{\Theta}$.}

The term $S_\mathrm{DM}(\mathrm{DM_{obs}})$ denotes the instrumental selection function of $\mathrm{DM_{obs}}$.
In the CHIME Catalog I, this function peaks at $\mathrm{DM_{obs}} \simeq 1000~\mathrm{pc~cm^{-3}}$ and declines toward both lower and higher DMs. 
This behavior arises primarily from two effects: at high DMs, DM smearing dilutes the burst signal in time, whereas at low DMs, the CHIME radio frequency interference (RFI) mitigation strategy removes a significant fraction of the signal from very bright low-DM events.
Similar to $S_F(F)$, we adopt an empirical form for $S_\mathrm{DM}(\mathrm{DM_{obs}})$, given by~\citep{2022MNRAS.511.1961H}:
 \begin{align}\label{eq:SDM}
 	S_\mathrm{DM}(\mathrm{DM_{obs}}) = & -0.7707 (\lg \mathrm{DM_{obs}})^2 \nonumber \\ 
        &+ 4.5601 (\lg \mathrm{DM_{obs}}) - 5.6291.
 \end{align}
 
Substituting the expressions derived above into Eq.~(\ref{eq:P_obs}), we obtain the PDF of $\mathrm{DM_{obs}}$, $P(\mathrm{DM_{obs}}|\bm{\Theta})$.
The likelihood for the unlocalized FRB sample is then given by
\begin{equation}
	\mathcal{L}(\mathrm{DM_{obs}}|\bm{\Theta})=\prod_{i=1} ^{N} P(\mathrm{DM}_{\mathrm{obs},i}|\bm{\Theta}),
 \end{equation}
where $N$ is the number of observed unlocalized FRBs.
{We note that this likelihood does not include the term depending on the total number of expected detections that appears in Eq.~(27) of \cite{2022MNRAS.509.4775J}. Such a term in principle carries cosmological information through the astrophysical event rate, but it can be removed by adopting an uninformative prior on the total number, as is standard practice in gravitational-wave cosmology~\citep{2018ApJ...863L..41F, 2019MNRAS.486.1086M, 2023AJ....166...22G}. In our analysis, each $P(\mathrm{DM}_{\mathrm{obs},i}|\bm{\Theta})$ in the likelihood is individually normalized through Eq.~(\ref{eq:P_obs}), so that only the shape of the DM distribution is used and this term can be safely omitted here.}
The parameter set $\bm{\Theta}$ includes: $H_0$, $\Omega_\mathrm{m0}$, $\Omega_\mathrm{b0}$, and $f_d$ in Eq.~(\ref{eq:DMcos}); the host galaxy parameters $\mu_{\mathrm{host}}$ and $\sigma_{\mathrm{host}}$ in Eq.~(\ref{eq:P_host}); the parameters $n$, $\gamma$ and  $E_*$ in Eqs.~(\ref{eq:Rz}) and (\ref{eq:PhiE}); the spectral index $\alpha$; and $\mathrm{DM_{halo}}$.
Here, $E_*$ is sampled in $\lg$ space, i.e., $\lg E_*$.
A complete summary of these parameters is provided in Tab.~\ref{tab:1}.
The posterior PDF for the model parameters $\bm{\Theta}$ is then derived via Bayes' theorem:
 \begin{equation}\label{eq:posterior}
 \mathcal{P}(\bm{\Theta}|\mathrm{DM_{obs}}) \propto \mathcal{L}(\mathrm{DM_{obs}}|\bm{\Theta}) \times \Pi(\bm{\Theta}),
 \end{equation}
 where $\Pi$ denotes the prior PDF of the parameters.
In our baseline analysis, we fix the background cosmology to $\Omega_\mathrm{m0}=0.315$, $\Omega_\mathrm{b0}h^2=0.02236$~\citep{2020A&A...641A...6P}, and set $f_d=0.94$~\citep{2025NatAs...9.1226C}.
The remaining free parameters are constrained simultaneously by sampling the posterior PDF with \texttt{emcee}~\citep{2013PASP..125..306F} and are marginalized over, except for $H_0$.
The priors on these free parameters are listed in Tab.~\ref{tab:1}.

\begin{table}[htb]
	\caption{\label{tab:1} Prior distributions and fixed parameter values adopted for the model parameters. The symbol $\mathcal{U}$ denotes a uniform prior.}
		\centering
		\begin{tabular}{l r}
			\hline
			Parameter ~&~ Prior/Fixed value \\
			\hline
			$H_0$ & $\mathcal{U}(30,200)$ \\
			$\Omega_{\mathrm{m0}}$ & $0.315$ \\
			$\Omega_{\mathrm{b0}}h^2$ & $0.02236$ \\
			$f_d$ & $0.94$ \\
			\hline
			$\mu_{\mathrm{host}}$ & $\mathcal{U}(1,5.5)$ \\
			$\sigma_{\mathrm{host}}$ & $\mathcal{U}(0.1,1.5)$ \\
			$\mathrm{DM}_{\mathrm{halo}}$ & $\mathcal{U}(0,60)$ \\
			\hline
			$n$ & $\mathcal{U}(0,2)$ \\
			$\gamma$ & $\mathcal{U}(-4,0)$ \\
			$\lg E_*$ & $\mathcal{U}(38,43)$ \\
			$\alpha$ & $\mathcal{U}(-4,2)$ \\
			\hline
		\end{tabular}
\end{table}

\section{Results}\label{sec:results}
\begin{figure}[tbp]
    \centering
    \includegraphics[width=0.49\textwidth]{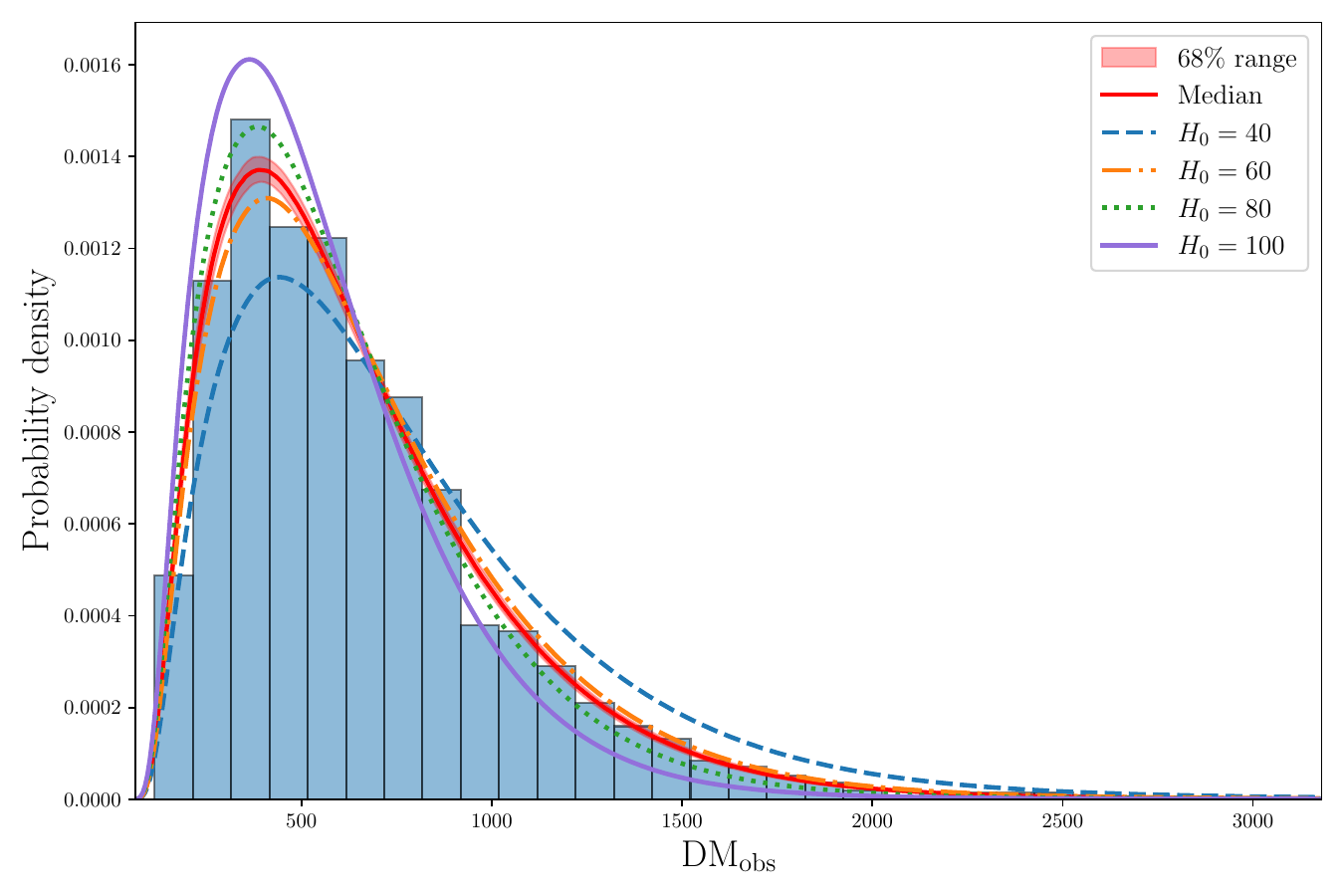}
    \caption{Comparison between the observed $\mathrm{DM_{obs}}$ distribution and the model
    inference for the 2124 unlocalized FRBs selected from the CHIME Catalog II. The histogram represents the data, whereas the red curve and the accompanying shaded band indicate the posterior median and the 68\% CL of the inferred DM distribution, as derived from the posterior samples of the model parameters. The additional curves show the theoretical $P(\mathrm{DM_{obs}}|\bm{\Theta})$ computed from Eq.~(\ref{eq:P_obs}) for several fixed values of $H_0$ (40, 60, 80, and $100~\kmsMpc$), with all other parameters fixed at their posterior median values. 
        \label{fig:DMobs}}
\end{figure}

The unlocalized FRB sample used in our analysis is taken from the latest released CHIME Catalog II,
which comprises 3641 unique sources detected between 25 July 2018 and 15 September 2023~\citep{2026ApJS..283...34C}. 
We adopt the $\mathrm{DM_{obs}}$ values derived using the \texttt{fitburst} routine. 
To ensure the reliability of our cosmological constraints, we select a subsample by retaining only events that satisfy all of the following criteria: (1) unlocalized, nonrepeating FRBs; (2) events with \texttt{excluded\_flag=0}; (3) events with a signal-to-noise ratio (SNR) greater than 10; (4) events with scattering timescales shorter than $10~\mathrm{ms}$; (5) events with $\mathrm{DM_{obs}}>1.5\,\mathrm{DM_{ISM}}$ to ensure an extragalactic origin, where $\mathrm{DM_{ISM}}$ is calculated using the YMW16 model~\citep{2017ApJ...835...29Y}; and (6) events with $\mathrm{DM_{obs}}-\mathrm{DM_{ISM}}>60~\pccm$, motivated by our assumption that $\mathrm{DM_{halo}}$ does not exceed this value. Applying these criteria yields a final sample of 2124 unlocalized FRBs.
{The above cuts are closely related to the selection functions adopted in our modeling. The empirical forms of $S_F$ and $S_\mathrm{DM}$ are taken from \cite{2022MNRAS.511.1961H}, whose construction adopted selection criteria similar to criteria (1)--(5) above. The additional cut is criterion (6), which we impose to ensure consistency with the adopted prior $\mathrm{DM_{halo}}\in\mathcal{U}(0,60)~\pccm$. This cut removes only three events with small $\mathrm{DM_{obs}}$ and does not alter the overall shape of the $\mathrm{DM_{obs}}$ distribution, so its impact on the cosmological constraints is negligible. Applying these cuts therefore introduces no significant unmodeled selection bias~\citep{2013NJPh...15e3027M}.}
Figure~\ref{fig:DMobs} compares the observed $\mathrm{DM_{obs}}$ distribution of the 2124 selected unlocalized FRBs with the corresponding model inference. The red curve and the shaded region indicate the median and the 68\% confidence level (CL) of the inferred distribution, respectively, as computed from Eq.~(\ref{eq:P_obs}) using parameter samples drawn from the posterior.
{To illustrate how the inferred distribution depends on $H_0$, Fig.~\ref{fig:DMobs} also shows the theoretical distributions obtained for several fixed values of $H_0$, with all other parameters fixed at their posterior median values. Since $\langle\mathrm{DM_{cos}}\rangle\propto H_0^{-1}$ and the scatter of $\mathrm{DM_{cos}}$ scales with its mean, a smaller $H_0$ shifts the distribution toward higher $\mathrm{DM_{obs}}$ and makes it broader with a lower peak, whereas a larger $H_0$ shifts the distribution toward lower $\mathrm{DM_{obs}}$ and makes it narrower and more sharply peaked. It is this sensitivity of the distribution shape to $H_0$ that allows the observed $\mathrm{DM_{obs}}$ distribution to constrain $H_0$.}
The overall consistency between the model and the data suggests that our adopted framework provides a statistically acceptable description of the observed DM distribution.

\begin{figure}[tbp]
	\centering
	\includegraphics[width=0.49\textwidth]{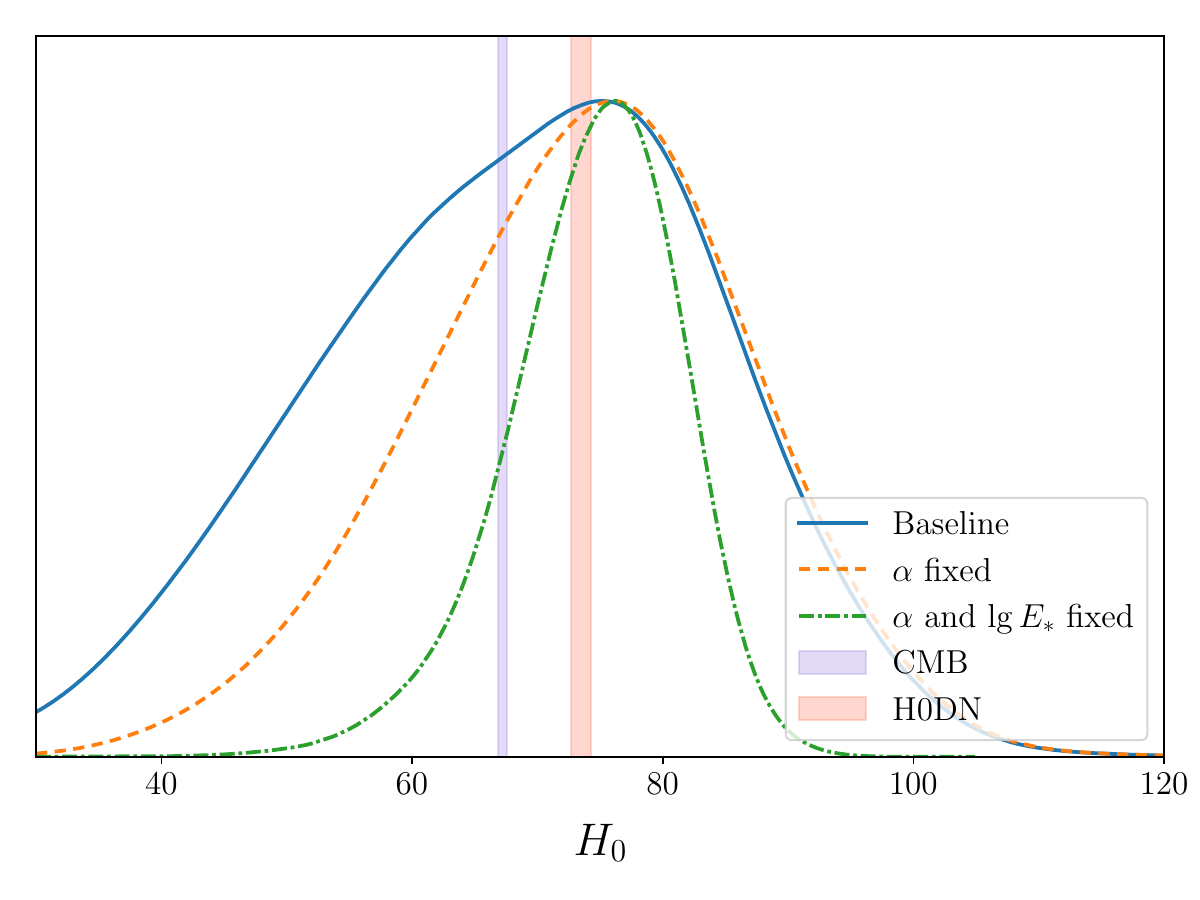}
	\caption{
		{Marginalized posterior distributions of $H_0$ from the unlocalized FRB sample: blue solid line for the baseline analysis, orange dashed line for $\alpha$ fixed at $-1.39$, and green dash-dotted line for both $\alpha$ and $\lg E_*$ fixed at $-1.39$ and $40.9$, respectively.}
		For comparison, the purple and red shaded bands represent the $1\sigma$ constraints from the latest CMB inference ($H_0=67.24\pm0.35~\kmsMpc$) and the local distance-indicator measurement ($H_0=73.50\pm0.81~\kmsMpc$), respectively.
			\label{fig:H0}}
\end{figure}

\begin{figure*}[tbp]
	\centering
	\includegraphics[width=0.8\textwidth]{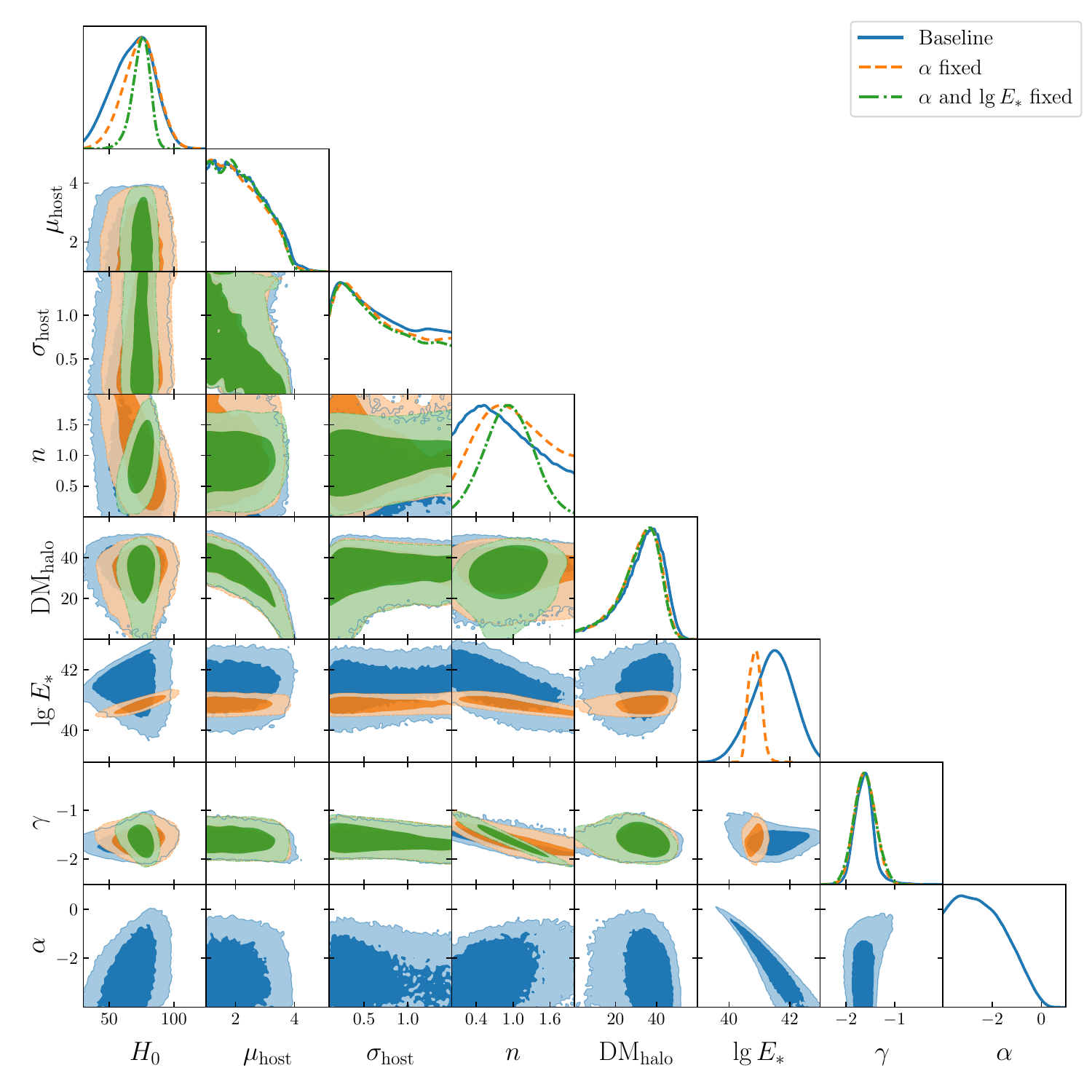}
	\caption{
		Posterior constraints on $H_0$, $\mu_{\mathrm{host}}$, $\sigma_{\mathrm{host}}$, $\mathrm{DM_{halo}}$, $n$, $\gamma$, $\lg E_*$, and $\alpha$ derived from 2124 unlocalized FRBs under three assumptions. Blue solid, orange dashed, and green dash-dotted contours correspond to the baseline analysis, $\alpha=-1.39$ fixed, and both $\alpha=-1.39$ and $\lg E_*=40.9$ fixed, respectively. The 1D posterior distributions and 2D confidence regions show the $1\sigma$ and $2\sigma$ confidence levels. 
		\label{fig:A_contour}}
\end{figure*}

The marginalized posterior distribution of $H_0$ is shown by the blue line in Fig.~\ref{fig:H0}, while those of the other parameters are presented in Fig.~\ref{fig:A_contour}. 
Using only the unlocalized FRB sample, our baseline analysis yields a constraint of $H_0 = 69^{+17}_{-15}~\kmsMpc$ at the 1$\sigma$ CL, with a standard deviation of $\sigma_{H_0}=15~\kmsMpc$. This corresponds to a relative uncertainty of approximately $\sigma_{H_0}/\langle H_0 \rangle \simeq 22\%$, where $\langle H_0 \rangle$ denotes the mean value of $H_0$. Figure~\ref{fig:H0} also shows the constraints from the latest CMB inference, $H_0=67.24\pm0.35~\kmsMpc$~\citep{2026PhRvD.113h3504C}, and from the latest local distance-indicator measurement by the H0DN collaboration, $H_0=73.50\pm0.81~\kmsMpc$~\citep{2026A&A...708A.166H}. Given the relatively large uncertainty, our result is consistent with both measurements at the $1\sigma$ CL.

The posterior contours presented in Fig.~\ref{fig:A_contour} reveal pronounced degeneracies among $H_0$, the spectral index $\alpha$, and the characteristic cutoff energy $\lg E_*$. This implies that the inferred value of $H_0$ is sensitive to both $\alpha$ and $\lg E_*$, indicating that the uncertainty on $H_0$ could be substantially reduced if these two parameters are independently constrained. 
To assess this improvement, we sequentially fix $\alpha$ and $\lg E_*$ in two separate steps. Since our current analysis yields only an upper limit on $\alpha$ (see Tab.~\ref{tab:A1}), we first adopt the value inferred from the CHIME Catalog I by \cite{2023ApJ...944..105S}, fixing $\alpha=-1.39$ while allowing all remaining parameters to vary freely. With this constraint, we obtain $H_0 = 74^{+14}_{-12}~\kmsMpc$, corresponding to $\sigma_{H_0}=13~\kmsMpc$ and a relative uncertainty of $18\%$ (shown as the orange dashed line in Fig.~\ref{fig:H0}). Compared with the baseline analysis, the uncertainty on $H_0$ is visibly reduced, as the degeneracy between $H_0$ and $\alpha$ is effectively broken. The remaining uncertainty is now dominated by the degeneracy between $H_0$ and $\lg E_*$, as shown by the orange contour in Fig.~\ref{fig:A_contour}.
To further quantify the improvement achievable when $\lg E_*$ is also determined, we additionally fix $\lg E_*$ to the mean value obtained from the $\alpha$-fixed analysis above, $\lg E_*=40.9$ (see Tab.~\ref{tab:A1}), while retaining $\alpha=-1.39$. Re-constraining the remaining parameters with the same 2124 FRBs yields $H_0 = 74^{+8}_{-6}~\kmsMpc$ with $\sigma_{H_0}=7~\kmsMpc$ (green dash-dotted line in Fig.~\ref{fig:H0}), reducing the relative uncertainty from $18\%$ to $9\%$. This central value is slightly higher than that inferred from the CMB, with a discrepancy of about $1.1\sigma$.

\begin{table}[tbp]
\caption{\label{tab:A1} 
Summary of marginalized parameter constraints for three different model setups: the baseline analysis (all parameters free), with $\alpha$ fixed, and with $\alpha$ and $\lg E_*$ jointly fixed. Values are given as posterior means with $1\sigma$ CL uncertainties.}
\centering
\begin{tabular}{l| c c c}
\hline
\hline
Parameter & \multicolumn{3}{c}{Constraint} \\
\hline
&  Baseline & $\alpha$ fixed & $\alpha$ and $\lg E_*$ fixed \\
\hline
$H_0$ & $69^{+17}_{-15}$ & $74^{+14}_{-12}$ & $74^{+8}_{-6}$ \\
$\mu_{\mathrm{host}}$ & $<2.6$ & $<2.5$ & $<2.6$ \\
$\sigma_{\mathrm{host}}$ & $<0.9$ & $<0.9$ & $<0.9$ \\
$\mathrm{DM}_{\mathrm{halo}}$ & $33^{+13}_{-6}$ & $31^{+12}_{-6}$ & $31^{+12}_{-6}$ \\
$n$ & $<1.1$ & $1.0^{+0.5}_{-0.7}$ & $0.9\pm0.4$ \\
$\gamma$ & $-1.6^{+0.2}_{-0.2}$ & $-1.6\pm0.2$ & $-1.6\pm0.2$ \\
$\lg E_*$ & $41.4^{+0.7}_{-0.6}$ & $40.9^{+0.2}_{-0.2}$ & $40.9$ (fixed)  \\
$\alpha$ & $<-1.9$ & $-1.39$ (fixed) & $-1.39$ (fixed) \\
\hline
\end{tabular}
\end{table}

\begin{figure*}[tbp]
	\centering
	\includegraphics[width=0.85\textwidth]{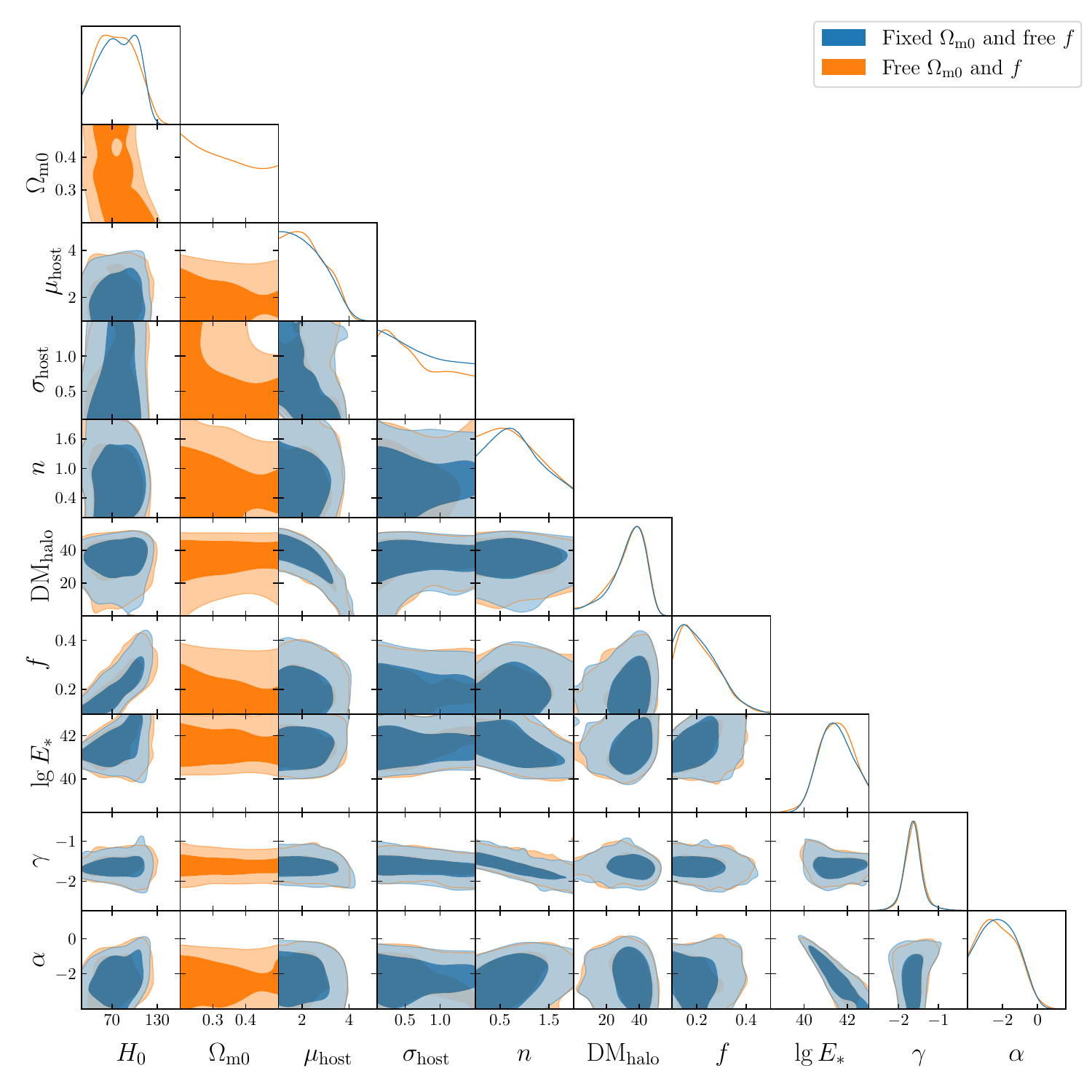}
	\caption{
		{\bf Posterior constraints from the two robustness tests. Blue contours correspond to the case with $\Omega_\mathrm{m0}=0.315$ fixed, and orange contours to the case with both $f$ and $\Omega_\mathrm{m0}$ free. 
		}
		\label{fig:triangle_omm_f}}
\end{figure*}

{Besides $\alpha$ and $\lg E_*$ discussed above, our baseline analysis also fixes two other parameters that are correlated with $H_0$ in the $\mathrm{DM_{obs}}$ distribution. 
One is the matter density parameter $\Omega_\mathrm{m0}$, which is degenerate with $H_0$ at high redshift through Eq.~(\ref{eq:DMcos}). 
The other is the parameter $f$ introduced in Sec.~\ref{sec:method}, which modifies the shape of $P_\mathrm{cos}$ and hence the $\mathrm{DM_{obs}}$ distribution on which our method relies~\citep{2024ApJ...965...57B, 2026ApJ...996...66Z}. 
While $\Omega_\mathrm{m0}$ is already tightly constrained by CMB and Type Ia supernova measurements~\citep{2020A&A...641A...6P, 2022ApJ...938..110B}, the true value of $f$ remains poorly known. 
If the adopted values of these two parameters deviate from the true ones, the inferred $H_0$ could be biased. 
Therefore, we test the robustness of the inferred $H_0$ by repeating the analysis in two cases, first freeing only $f$ under a uniform prior $f\in\mathcal{U}(0.1, 0.5)$ while keeping $\Omega_\mathrm{m0}=0.315$ fixed, and then freeing both $f$ and $\Omega_\mathrm{m0}$ with $\Omega_\mathrm{m0}\in\mathcal{U}(0.2,0.5)$. 
As shown in Fig.~\ref{fig:triangle_omm_f}, $f$ is positively correlated with $H_0$, whereas $\Omega_\mathrm{m0}$ is mildly anti-correlated with $H_0$ and is unconstrained. This is expected since $\langle\mathrm{DM_{cos}}\rangle$ depends on $\Omega_\mathrm{m0}$ more weakly than on $H_0$ in Eq.~(\ref{eq:DMcos}). 
In the first case we obtain an upper limit of $f<0.25$ at the 68\% CL, and in the second case, $f=0.22^{+0.03}_{-0.11}$. Both results are consistent with the baseline value $f=10^{-0.75}\simeq0.178$. 
The inferred $H_0$ shifts mildly higher than the baseline result, to $H_0 = 78^{+31}_{-22}~\kmsMpc$ and $H_0 = 77^{+24}_{-29}~\kmsMpc$ in the two cases, respectively. Their uncertainties, however, become larger than those of the baseline analysis, so both results remain consistent with the baseline value $H_0 = 69^{+17}_{-15}~\kmsMpc$ within $1\sigma$. 
The other population parameters are also consistent with the baseline results in both cases.
These results indicate that, at the current statistical precision, fixing $f$ and $\Omega_\mathrm{m0}$ in the baseline analysis does not introduce a significant bias on the inferred $H_0$.
}

To further assess the precision achievable with this method, we generate mock catalogs of unlocalized FRBs and calculate the relative uncertainty in the inferred $H_0$. The fiducial model used to construct the mock sample adopts $H_0=70~\kmsMpc$, $\Omega_{\mathrm{m0}}=0.315$, $\Omega_{\mathrm{b0}}h^2=0.02236$, $f_d=0.94$, $\mu_{\mathrm{host}}=4$, $\sigma_{\mathrm{host}}=0.5$, $n=1$, $\gamma=-1.3$, $\lg E_*=41.38$, $\alpha=-1.39$, and $\mathrm{DM_{MW}}=100~\pccm$.
The mock $\mathrm{DM_{ext}}$ values are sampled according to Eq.~(\ref{eq:P_obs}). 
With these mock catalogs, we test the $H_0$ constraint precision under the same two assumptions as for the real data: with $\alpha$ fixed, and with both $\alpha$ and $\lg E_*$ fixed.
In the first case, fixing $\alpha$ and leaving $\mathrm{DM_{halo}}$ unconstrained, we infer $H_0 = 70^{+6}_{-5}~\kmsMpc$ ($\sigma_{H_0}=5~\kmsMpc$) for a catalog of 5000 FRBs, corresponding to a relative uncertainty of approximately 7\%.
In the second case, with $\lg E_*$ additionally fixed to its fiducial value, the relative uncertainty is further reduced to approximately 3\% (with $\sigma_{H_0}\simeq2~\kmsMpc$).
{At this precision, the latest CMB-inferred and local distance-indicator values of $H_0$, which differ by $\simeq 6~\kmsMpc$~\citep{2026PhRvD.113h3504C, 2026A&A...708A.166H}, could already be distinguished at the $\sim3\sigma$ CL using unlocalized FRBs alone. If the mock sample is further increased to 30000 FRBs with the same fixed $\lg E_*$, the relative uncertainty in $H_0$ can be reduced to nearly $1\%$ ($\sigma_{H_0}\simeq1~\kmsMpc$), corresponding to a discrimination between the two values at more than $5\sigma$.}
These tests demonstrate that the constraining power of this method can be substantially enhanced with future larger samples of unlocalized FRBs, particularly if the FRB energy distribution can be independently calibrated with higher precision.

\section{Conclusion}\label{sec:conclusion}
In summary, we have demonstrated that the observed DM distribution of unlocalized FRBs can already provide an effective constraint on $H_0$. In our baseline analysis, the unlocalized FRB sample alone yields a measurement of $H_0 = 69^{+17}_{-15}~\kmsMpc$, corresponding to an uncertainty of about 22\%. This constitutes the first independent $H_0$ measurement derived solely from the DM distribution of unlocalized FRBs, offering a new avenue to address the Hubble tension. 
Our analysis also highlights the critical role of calibrating the FRB population parameters. Fixing both the spectral index $\alpha$ and the characteristic cutoff energy $\lg E_*$ effectively breaks the dominant degeneracies and reduces the $H_0$ uncertainty to 9\%, thereby demonstrating a clear path toward competitive precision.
Mock catalog tests further show that, with a sample of 5000 FRBs, this uncertainty can be reduced to approximately 3\%. {At this precision, the CMB-inferred and local distance-indicator values of $H_0$ could already be distinguished at the $\sim3\sigma$ level using unlocalized FRBs alone, and increasing the sample to 30000 FRBs could further reduce the uncertainty to nearly $1\%$ and distinguish the two values at more than $5\sigma$.}
{The sample required by our method is expected to grow rapidly. With CHIME/FRB continuing to operate and future survey telescopes such as DSA-2000 designed to detect and localize $\sim10^4$ FRBs per year~\citep{2019BAAS...51g.255H}, a sample of 5000 unlocalized FRBs is expected to become available in a few years, without requiring any host galaxy identification or spectroscopic follow-up.
In comparison, reaching $\sim1\%$ precision with localized FRBs requires about 500 sources with redshift measurements~\citep{2025A&A...698A.215G}. Although a sample of more than 1000 precisely localized FRBs is among the key survey requirements of the CHIME/FRB Outriggers~\citep{2025ApJ...993...55C}, the growth of samples with measured redshifts remains limited by host galaxy identification and spectroscopic follow-up rather than by localization alone.}
Our analyses establish unlocalized FRBs as a promising and complementary probe for future precision measurements of the Hubble constant.

\section*{Acknowledgments}
This work is supported by the National Natural Science Foundation of China (grant Nos. 12603008, 12422307, 12373053, 12321003, and 12275080), 
the Strategic Priority Research Program of the Chinese Academy of Sciences (grant No. XDB0550400), the National Key R\&D 
Program of China (grant Nos. 2024YFA1611704 and 2024YFA1611700), the China Manned Space Program (grant No. CMS-CSST-2025-A01), and China Postdoctoral Science Foundation (grant No. 2025M783235).

\bibliographystyle{aasjournalv7}
\bibliography{ref_ADS}

\end{document}